\documentclass[11pt,a4paper]{article}
\usepackage{fancyhdr}
\usepackage{palatino}
\usepackage{amssymb}
\usepackage{epsfig}
\usepackage{vmargin}
\usepackage{natbib}

\makeatletter \renewcommand{\@biblabel}[1]{#1.} \makeatother

\begin{document}

\begin{bfseries}
\noindent\Large{Pitch perception: A dynamical-systems perspective}
\vspace{0.5cm}

\noindent\small{Julyan H. E. Cartwright*, Diego L. Gonz\'alez\dag,
 \& Oreste Piro\ddag}
\end{bfseries}

\begin{itshape}
\noindent\footnotesize
*~Laboratorio de Estudios Cristalogr\'aficos, CSIC, E-18071 Granada, Spain. \\
E-mail julyan@lec.ugr.es,  Web http://lec.ugr.es/$\sim$julyan \\
\dag~Istituto Lamel, CNR, I-40129 Bologna, Italy. \\
E-mail gonzalez@lamel.bo.cnr.it \\
\ddag~Institut Mediterrani d'Estudis Avan\c{c}ats, CSIC--UIB, 
E-07071 Palma de Mallorca, Spain. \\
E-mail piro@imedea.uib.es, Web http://www.imedea.uib.es/$\sim$piro \\
\end{itshape}

\normalsize

Published in: Proc. Nat. Acad. Sci. USA {\bf 98}, 4855-4859, 2001.

\vspace{0.5cm}





\begin{bfseries}
\noindent\small
Two and a half millennia ago Pythagoras initiated the scientific study of the
pitch of sounds; yet our understanding of the mechanisms of pitch perception
remains incomplete. Physical models of pitch perception try to explain from
elementary principles why certain physical characteristics of the stimulus lead
to particular pitch sensations. There are two broad categories of
pitch-perception models: place or spectral models consider that pitch is mainly
related to the Fourier spectrum of the stimulus, whereas for periodicity or
temporal models its characteristics in the time domain are more important.
Current models from either class are usually computationally intensive,
implementing a series of steps more or less supported by auditory physiology.
However, the brain has to analyse and react in real time to an enormous amount
of information from the ear and other senses. How is all this information
efficiently represented and processed in the nervous system? A proposal of
nonlinear and complex systems research is that dynamical attractors may form
the basis of neural information processing. Because the auditory system is a 
complex and highly nonlinear dynamical system it is natural to suppose that
dynamical attractors may carry perceptual and functional meaning. Here we show
that this idea, scarcely developed in current pitch models, can be successfully
applied to pitch perception.
\end{bfseries}

\begin{figure}[p]
\begin{center}
\epsfig{file=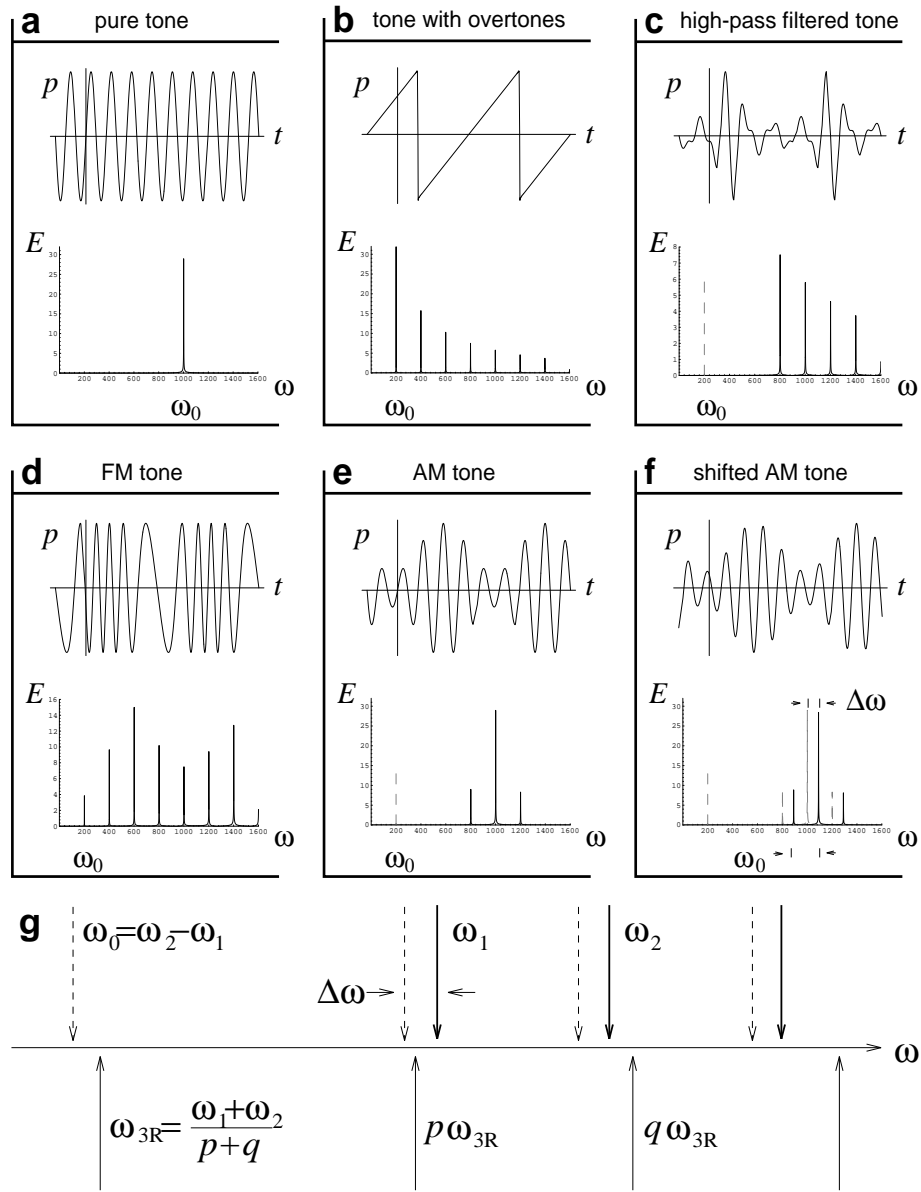,width=0.8\textwidth}
\end{center}
\caption{\label{stimuli}
Stimuli: waveforms, Fourier spectra, and pitches.
(a) 1 kHz pure tone; the pitch coincides with the frequency $\omega_0$.
(b) Complex tone formed by 200 Hz fundamental plus overtones; the pitch is at
the frequency of the fundamental $\omega_0$.
(c) After high-pass filtering of the previous tone to remove
the fundamental and the first few overtones, the pitch $\omega_0$ remains at the
frequency of the missing fundamental (dotted).
(d) The result of frequency modulation of a 1 kHz pure tone carrier
by a 200 Hz pure tone modulant.
(e) Complex tone produced by amplitude modulation of a 1 kHz pure tone carrier
by a 200 Hz pure tone modulant; the pitch coincides with the difference 
combination tone $\omega_0$.
(f) Result of shifting the partials of the previous tone in frequency by 
$\Delta\omega=90$ Hz; the pitch shifts by $\Delta\omega_0\approx 20$ Hz,
although the difference combination tone does not.
(g) Schematic diagram of the frequency line details (above the line)
the pitch shift behaviour of (f) and (below the line) the three-frequency 
resonance we propose to explain it.
}
\end{figure}

The pitch of a sound is where we perceive it to lie on a musical scale. For a
pure tone with a single frequency component, pitch rises monotonically with
frequency. However, more complex signals also elicit a pitch
sensation. Some instances are presented in Fig.~\ref{stimuli}. These are sounds
produced by the nonlinear interaction of two or more periodic sources, by
amplitude or frequency modulation. All such stimuli, which may be termed
complex tones, produce a definite pitch sensation, and all of them exhibit a
certain spectral periodicity. Many natural sounds have this quality,
including vowel sounds in human speech and vocalizations of many other
animals. Evidence for the importance of spectral periodicity in sound
processing by humans is that noisy stimuli exhibiting this property also
elicit a pitch sensation. An example is repetition pitch: the pitch of 
ripple noise \cite{yost}, which arises naturally when the sound from a noisy
source interacts with a delayed version of itself, produced, for example, by
a single or multiple echo. It is clear that an efficient mechanism for the
analysis and recognition of complex tones represents an evolutionary advantage
for an organism. In this light, the pitch percept may be seen as an effective
one-parameter categorization of sounds possessing some spectral periodicity
\cite{bregman,hartmann,roberts,moore}. 

\begin{center}\bf Virtual Pitch\end{center}

For a harmonic stimulus like Fig.~\ref{stimuli}b (a periodic signal), 
there is a natural physical 
solution to the problem of encoding it with a single parameter: take the
fundamental component of the stimulus as the pitch and all other components are
naturally recorded as the higher harmonics of the fundamental. This is what 
nature does. However, a harmonic stimulus like Fig.~\ref{stimuli}c, which is 
high-pass filtered so that the fundamental and some of the first higher
harmonics are eliminated, nevertheless maintains its pitch at the frequency of
the absent fundamental. The stimulus (Fig.~\ref{stimuli}e) obtained by
amplitude modulation of a sinusoidal carrier of 1 kHz by a sinusoidal modulant
of 200 Hz is also of this type. As the carrier and modulant are rationally
related, the stimulus is harmonic; the partials are integer multiples of the
absent fundamental $\omega_0=200$ Hz. The perception of pitch for this kind of
stimulus is known as the problem of the missing fundamental, virtual pitch, or
residue perception \cite{deboer}. The first physical theory  for the phenomenon
was proposed by von Helmholtz \cite{helmholtz}, who attributed it to the
generation of difference combination tones in the nonlinearities of the ear. A
passive nonlinearity fed by two sources with frequencies $\omega_1$ and
$\omega_2$ generates combination tones of frequency $\omega_C$ (see the 
Appendix for clarification of the concepts from nonlinear dynamics used
throughout this paper). For a harmonic complex tone, such as
Fig.~\ref{stimuli}e, the difference combination tone
$\omega_C=\omega_2-\omega_1$ between two successive partials has the frequency
of the missing fundamental $\omega_0$. In a crucial experiment, however,
Schouten et al.\ \cite{schouten} demonstrated that the  residue cannot be
described by a difference combination tone: if we shift all the partials in
frequency by the same amount $\Delta\omega$  (Fig.~\ref{stimuli}f), the
difference combination tone remains unchanged. But the perceived pitch shifts,
with a linear dependence on $\Delta\omega$.

\begin{center}\bf A Dynamical-Systems Perspective\end{center}

Such a complex tone is no longer harmonic. How does nature encode an inharmonic
complex tone into a single pitch? Intuitively, the shifted pseudofundamental
depicted in Fig.~\ref{stimuli}g might seem to be a better choice than the
unshifted fundamental, which corresponds to the difference combination tone.
However, from a mathematical point of view, this is not obvious. The ratios
between successive partials of the shifted stimulus are irrational and we
cannot represent them as higher harmonics of a nonzero fundamental frequency; 
the true fundamental would have frequency zero. Some kind of
approximation is needed. The approximation of two arbitrary frequencies,
$\omega_1$ and $\omega_2$, by the harmonics of a third, $\omega_R$, is
equivalent to the mathematical problem of finding a strongly convergent
sequence of pairs of rational numbers with the same denominator that
simultaneously approximates the two frequency ratios, $\omega_1/\omega_R$ and
$\omega_2/\omega_R$. If we consider the approximation to only one frequency
ratio there exists a general solution given by the continued-fraction algorithm
\cite{kinchin}. However, for two frequency ratios a general solution is not
known. Some algorithms have been proposed that work for particular values
of the frequency ratios or that are weakly convergent \cite{kim1}. We developed
an alternative approach \cite{3freq}. The idea is to equate the distances
between appropriate harmonics of the pseudofundamental and the pair of
frequencies we wish to approximate. In this way the two approximations are
equally good or bad. The problem can then be solved by a generalization of the
Farey sum. This approach enables the hierarchical
classification of a type of dynamical attractors found in systems with three
frequencies: three-frequency resonances $[p,q,r]$.  

A classification of three-frequency resonances allows us to propose how nature
might encode an inharmonic complex tone into a single pitch percept. 
The pitch of a complex tone corresponds to a one-parameter
categorization of sounds by a physical frequency whose harmonics are good
approximations to the partials of the complex. This physical frequency is
naturally generated as a universal response of a nonlinear dynamical system 
--- the auditory system, or some specialized subsystem of it --- 
under the action of an external force, namely the
stimulus. Psychophysical experiments with multicomponent stimuli suggest that
the lowest-frequency components are usually dominant in determining residue
perception \cite{deboer}. Thus we represent the external force as a first
approximation by the two lowest-frequency components of the stimulus. For pitch
shift experiments with small frequency detuning $\Delta\omega$, such as those
of Schouten et al., the vicinity of these two lowest components
$\omega_1=k\omega_0+\Delta\omega$ and $\omega_2=(k+1)\omega_0+\Delta\omega$ to
successive multiples of some missing fundamental ensures that $(k+1)/k$ is a
good rational approximation to their frequency ratio. Hence we concentrate on a
small interval between the frequencies $\omega_1/k$ and $\omega_2/(k+1)$ around
the missing fundamental of the nonshifted case. These frequencies correspond to
the three-frequency resonances $[0,-1,k]$ and $[-1,0,k+1]$. We suppose that the
residue should be associated with the largest three-frequency resonance in this
interval: the daughter of these resonances, $[-1,-1,2k+1]$. If this reasoning is
correct, the three-frequency resonance formed between the two lowest-frequency
components of the complex tone and the response frequency
$P=(\omega_1+\omega_2)/(2k+1)$ gives rise to the perceived residue pitch $P$. 

\begin{center}\bf Results\end{center}

\begin{figure}[p]
\begin{center}
\epsfig{file=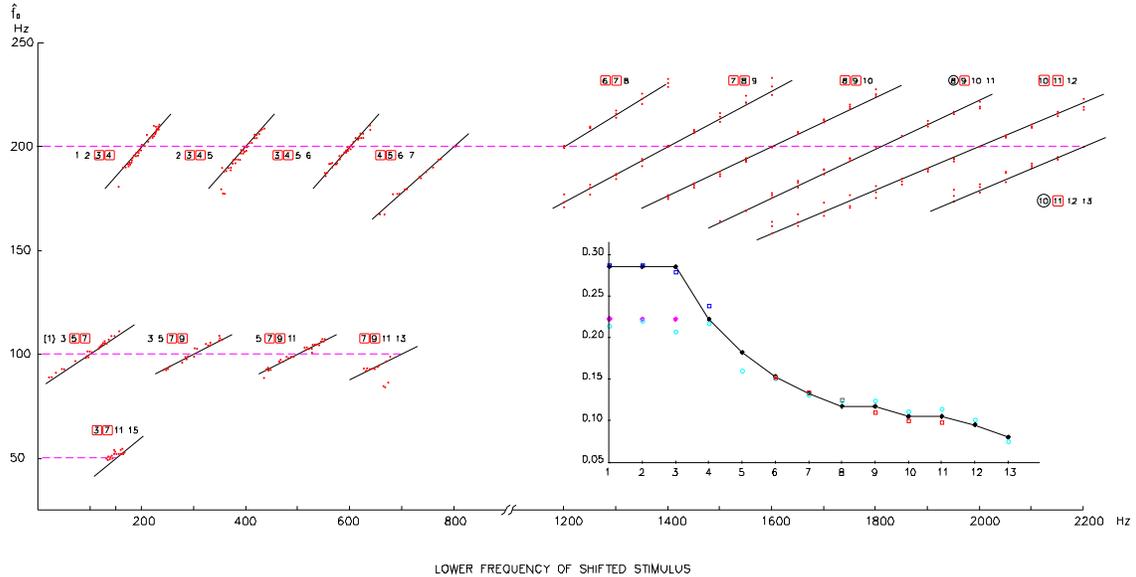,width=\textwidth}
\end{center}
\caption{\label{results}
Experimental data (red dots) from Gerson \& Goldstein \cite{gerson} (0--800 Hz
range) and from Schouten et al.\ \cite{schouten} (1200--2200 Hz range) show 
pitch as a function of the lower frequency $f=k\omega_0+\Delta\omega$ of a 
complex tone $\{k\omega_0+\Delta\omega, (k+1)\omega_0+\Delta\omega, 
(k+2)\omega_0+\Delta\omega, \ldots\}$ with the partials spaced
$g=\omega_0=200$  Hz apart. The data of Schouten et al.\ are for
three-component tones monotically presented (all of the stimulus entering one
ear), and those of Gerson \& Goldstein for four-component tones dichotically
presented (part of the stimulus entering one ear and the rest of the stimulus
the other, controlateral, ear); the harmonic numbers of the partials present in
the stimuli are shown beside the data. The pitch-shift effect we predict from
three-frequency resonance,
taking into account the dominance region, is shown superimposed on the data as
solid lines given by the equations $P=g+(f-n\,g)/(n+1/2)$ (primary lines)
$P=g/2+(f-(n+1/2)g)/(2n+2)$ (secondary lines), and $P=g/4+(f-(n-1/4)g)/(4n+1)$ 
(tertiary line); the harmonic numbers of the partials used to 
calculate the pitch-shift lines are shown enclosed in red squares. For primary
lines these harmonic numbers correspond to $n$ and $n+1$, for secondary lines to
$2n+1$ and $2n+3$, and for the tertiary line to $4n+1$ and $4n+5$. A red
circle, instead of a square, signifies that the component is not physically
present in the stimulus, but corresponds to a combination tone. The inset at
bottom right corresponds to the slopes of the data averaged  over the distinct
experimental values plotted as a function of harmonic number. The blue squares
are the data of Gerson \& Goldstein, the red squares are those of Schouten, and
lastly, the blue circles are data of Patterson \cite{patterson} for six and
twelve-component tones which are averaged over different experimental
situations that represent several thousand points. The black diamonds
correspond to our theory and show that the data of Gerson \& Goldstein and
those of Patterson saturate for different values of $k$ (the experimental
conditions were different). 
}
\end{figure}

As we showed in earlier work \cite{soundsletterprl}, there is good agreement
betwen the pitch perceived in experiments and the three-frequency resonance
produced by the two lowest-frequency components of the complex tone for
intermediate harmonic numbers $3\leq k\leq 8$. For high and low $k$ values
there are systematic deviations from these predictions. Such deviations, noted
in pitch-perception modelling, are explained by the dominance effect: there is
a frequency window of preferred stimulus components, so that not all components
are equally important in 
determining residue perception \cite{patterson2}. In order to describe these 
slope deviations for high and low $k$ values within our approach, we must,
instead of taking the lowest-frequency components, use some effective $k$ 
that depends on the dominance effect. In this, we also take into account the
presence of difference combination tones, which provide some components with
$k$s not present in the original stimulus.  In Fig.~\ref{results} we have
superimposed the predicted three-frequency resonances, including the dominance
effect, on published experimental pitch-shift data
\cite{schouten,gerson,patterson}. For stimuli consisting only of high-$k$
components, the window of the dominance region is almost empty, and difference
combination tones of lower $k$ can become more important than the primary
components in determining the pitch of the stimulus. The result of this
modification is a saturation of the slopes that correctly describes the
experimental data. A saturation of slopes can also be seen in the experimental
data for low values of $k$. This effect too can be explained in terms of the
dominance region. For a 200 Hz stimulus spacing, the region is situated at
about 800 Hz; this implies that stimulus components with harmonic numbers $n$
and $n+1$ other than the two lowest ones (i.e., $n>k$) become more important
for determining the the three-frequency resonance that provides the residue
pitch. Again, incorporating this modification, we can correctly predict
the experimental data.

But for the more complex case of low-$k$ stimuli, not only quantitative, but 
also qualitative, differences arise between the two-lowest-component theory and
experiment. The most interesting feature seen in the data of Fig.~\ref{results}
is a second series of pitch-shift  lines clustered around the pitch of 100 Hz. 
This too can be explained within the framework of our ideas. Recall that for
small frequency detuning, the frequency ratio between adjacent stimulus
components, $\Delta\omega$, can be approximated  by the quotient of two
integers differing by unity: $\omega_2/\omega_1=(n+1)/n$. However, if we relax
the small detuning constraint, so that $\Delta\omega$ becomes large, we can
move to a case where $\omega_2/\omega_1$ can better be approximated by
$(n+2)/(n+1)$. But, by the usual Farey sum operation between rational numbers,
we know that there exists between these two regions an interval in which the
frequency ratio can be better approximated by $(2n+3)/(2n+1)$. In this
interval, then, the main three-frequency resonance is $[-1,-1,4n+4]$, giving a
response frequency $P=(\omega_1+\omega_2)/(4n+4)$, which produces a pitch-shift
line with slope $1/(2n+2)$ around $\omega_0/2=100$ Hz for the case analysed. Of
course, if prefiltering produces a saturation of the slopes of the primary
pitch-shift lines, the same should occur for these secondary ones. In
Fig.~\ref{results} we show our predictions for the secondary lines taking in
account the dominance effect. The agreement, both qualitative and also
quantitative, is impressive. Moreover, a small group of data points indicates
the existence of a further level of pitch-shift lines clustered around 50 Hz
in a region between a primary and a secondary pitch-shift line. We can
understand this level in the same way as above, and we plot our
prediction for its pitch-shift line in Fig.~\ref{results}. This 
hierarchical arrangement of the perception of pitch of complex tones is
entirely consistent with the universal devil's staircase structure that
dynamical systems theory predicts for the three-frequency resonances in
quasiperiodically forced dynamical systems. Further evidence comes from
psychophysical experiments with pure tones. These, presented under particular
experimental conditions, also elicit a residue sensation. The extremes of the
three-frequency staircase correspond to subharmonics of only one external
frequency, and thus these are the expected responses when only one stimulus
component is present. As the results of Houtgast \cite{houtgast} show, these
subharmonics are indeed perceived. 

\begin{center}\bf Discussion\end{center}

A dynamical attractor can be studied by means of time or frequency analysis.
Both are common techniques in dynamical-systems analysis, but one is not
inherently more fundamental than the other, nor are these the only two tools
available. For this reason, and because our reasoning makes no use of a
particular physiological implementation, our results cannot be included
directly either in the spectral \cite{cohen2} or the temporal \cite{meddis}
classes of models of pitch perception. What we have proposed is not a model,
but a mathematical basis for the perception of pitch that uses the universality
of responses of dynamical systems to address the question of why the auditory
system should behave as it does when confronted by stimuli consisting of
complex tones. Not all pitch perception phenomena are explicable in terms of
universality; nor should they be, since some will depend on the specific
details of the neural circuitry. However, this is a powerful way of approaching
the problem that is capable of explaining many experimental data considered
difficult to understand. Future pitch models can surely incorporate these
results in their frameworks. Spectral models \cite{cohen2} can use these ideas
since they make consistent use of different kinds of harmonic templates, and
three-frequency resonances offer in a natural way optimized candidates for the
base frequency of such templates without the need to include stochastic terms.
Temporal models \cite{meddis} can apply these results as they need some kind of
locking of neural spiking to the fine structure of the stimulus, and
three-frequency resonances are the natural extension of phase locking to the
more complicated case of quasiperiodic forcing that is typically related to the
perception of complex tones. A dynamical-systems viewpoint can then integrate
spectral and temporal hypotheses into a coherent unified approach to pitch
perception incorporating both sets of ideas.

We have shown that universal properties of dynamical responses in nonlinear 
systems are reflected in the pitch perception of complex tones. In previous
work \cite{soundsletterprl}, we argued that a dynamical-systems approach backs
up experimental evidence for subcortical pitch processing in humans 
\cite{pantev}. The experimental evidence is not conclusive, as studies with
monkeys have found that raw spectral information is present in the primary
auditory cortex \cite{fishman}. However, whether this processing occurs in, 
or before, the auditory cortex, the dynamical mechanism we envisage greatly
facilitates processing of information into a single percept. Pitch processing
may then prove to be an example in which universality in nonlinear
dynamics can help to explain complex experimental results in biology. The
auditory system possesses an astonishing capability for processing 
pitch-related information in real time; what we have demonstrated here is how, 
at a fundamental level, this can be so.

\begin{center}\bf Acknowledgements\end{center}
We should like to thank Fernando Acosta for his help in the preparation of 
Fig.~\ref{results}. JHEC acknowledges the financial support of the Spanish
CSIC, and Plan Nacional del Espacio contract ESP98-1347. OP acknowledges the
Spanish Ministerio de Ciencia y Tecnologia, Proyecto CONOCE, contract
BFM2000-1108. DLG conceived the idea, and together with JHEC and OP carried 
out the research; JHEC and DLG cowrote the paper.

\begin{center}\bf References\end{center}

\bibliographystyle{nature}
\bibliography{database}

\newpage
\vspace*{0.5cm}
\begin{center}\bf Appendix: Universality in Nonlinear Systems\end{center}
\vspace{1cm}
Nonlinear systems exhibit universal responses under external forcing:
\vspace{1cm}
\subsubsection*{Harmonics from periodically forced passive nonlinearities}
\begin{center}
\epsfig{file=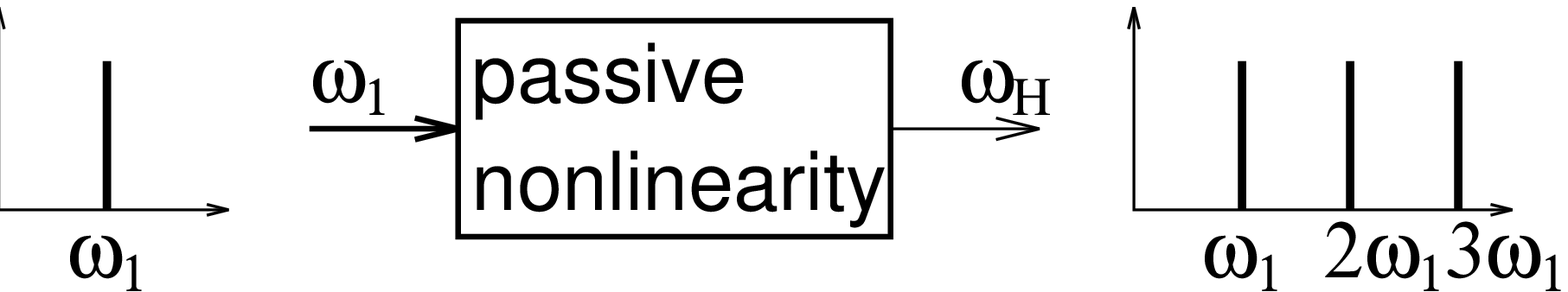,height=0.85cm}
\end{center}
A single frequency periodically forcing a passive (sometimes termed static) 
nonlinearity generates higher harmonics (overtones)
$2\omega_1, 3\omega_1, \ldots$
of a fundamental $\omega_1$, given by
$p\omega_1+\omega_{H}=0$ with $p$ integer. This is seen in acoustics
as harmonic distortion.
\vspace{1cm}
\subsubsection*{Combination tones from quasiperiodically forced passive 
nonlinearities}
\begin{center}
\epsfig{file=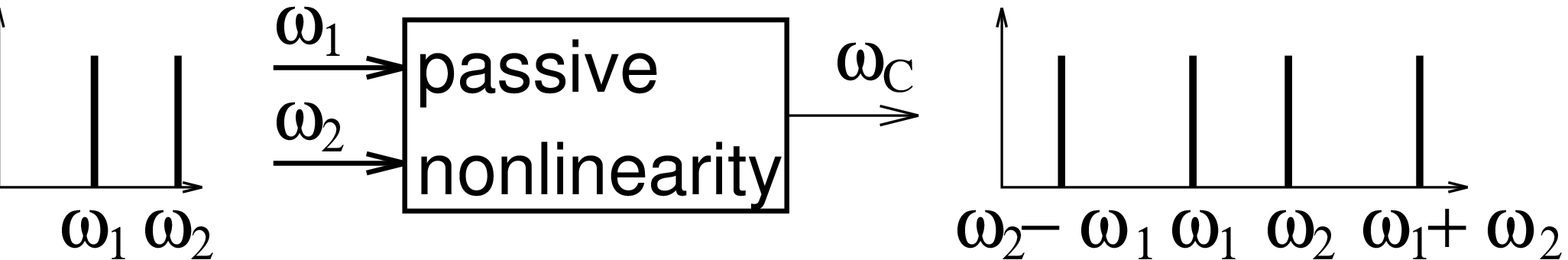,height=1cm}
\end{center}
A passive nonlinearity forced quasiperiodically by two sources generates
combination tones $\omega_1-\omega_2, \omega_1+\omega_2, \ldots$, which are 
solutions of the equation
$p\omega_1+q\omega_2+\omega_{C}=0$ where $p$ and $q$ are integers. They are
found as distortion products in acoustics. 
\vspace{1cm}
\subsubsection*{Subharmonics, or two-frequency resonances from periodically 
forced dynamical systems}
\begin{figure}[h]
\begin{center}
\epsfig{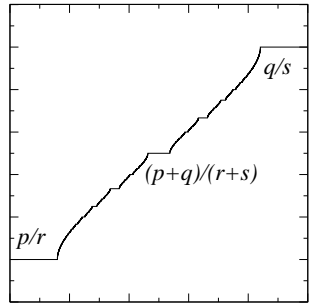}\end{center}
\caption{\label{devils_staircase}
Two-frequency devil's staircase.
The rotation number, the frequency 
ratio $\rho=-p/r=\omega_{2R}/\omega_1$, is
plotted against the period of the external force.
}
\end{figure}

\begin{center}
\epsfig{file=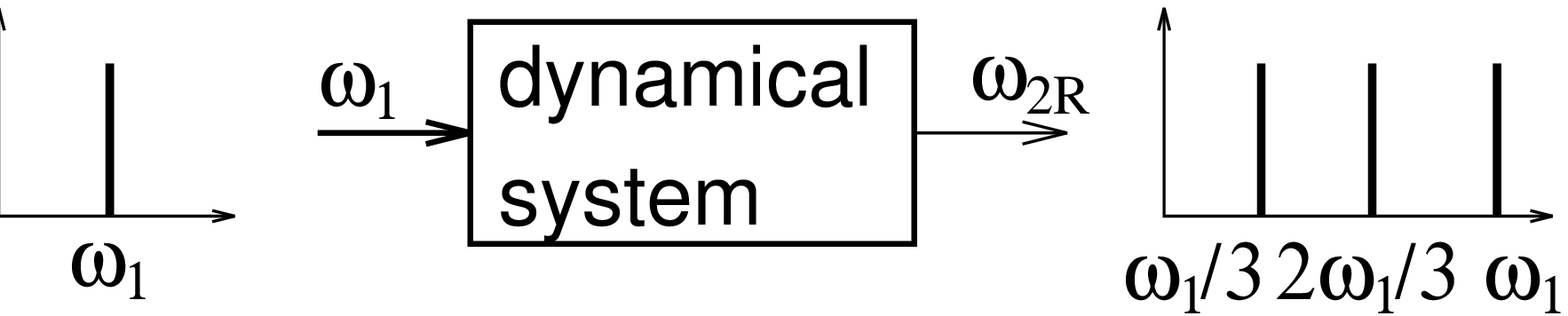,height=0.85cm}
\end{center}
With a periodically forced active nonlinearity --- a dynamical system --- more
complex subharmonic responses $\omega_1/r, 2\omega_1/r, \ldots, (r-1)\omega_1/r$
known as mode lockings or two-frequency
resonances are generated. These are given by $p\omega_1+r\omega_{2R}=0$ when
$p$ and $r$ are integers. 
As some parameter is varied, different resonances are found that remain
stable over an interval. A classical representation of this, known as
the devil's staircase, is shown in Fig.~\ref{devils_staircase}.

We see that the resonances are hierarchically arranged. The local ordering can
be described by the Farey sum: 
If two rational numbers $a/c$ and $b/d$ satisfy $|ad-bc|=1$ we say that they
are unimodular or adjacents and we can find between them a unique rational with
minimal denominator. This rational is called the mediant and can be expressed
as a Farey sum operation $a/c\oplus b/d=(a+b)/(c+d)$. The resonance
characterized by the mediant is the widest between those represented by the
adjacents \cite{oreste}.
\vspace{1cm}
\subsubsection*{Three-frequency resonances from quasiperiodically 
forced dynamical systems}
\begin{figure}[h]
\begin{center}
\epsfig{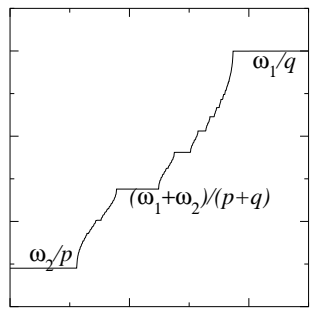}\end{center}
\caption{\label{3f_devils_staircase}
Three-frequency devil's staircase.
Contrarily to the case
of periodically driven systems, where plateaux represent periodic
solutions, here they represent quasiperiodic solutions
(only the third frequency is represented in the ordinate).
We have investigated these properties in three different systems: 
the quasiperiodic circle map, a system of coupled electronic oscillators 
and a set of ordinary nonlinear differential equations, with the same 
qualitative results \cite{3freqpaper}, which confirm the theoretical 
predictions \cite{3freq}.
}
\end{figure}

\begin{center}
\epsfig{file=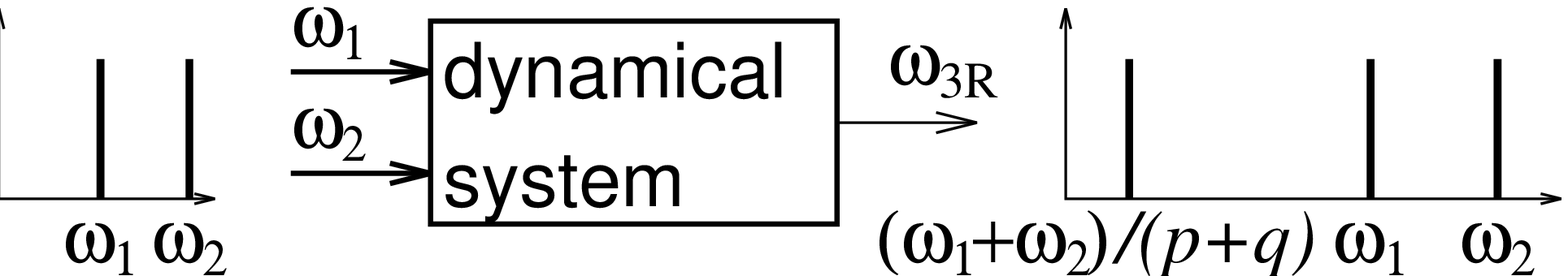,height=1cm}
\end{center}
Quasiperiodically forced dynamical systems show a great variety of
qualitative responses that fall into three main categories: there are 
periodic attractors, quasiperiodic attractors, and chaotic and nonchaotic 
strange attractors. Here we concentrate on the
three-frequency resonances produced by two-frequency quasiperiodic attractors 
as the natural candidates for modelling the residue \cite{lectnotes}. 
Three-frequency resonances are given by the nontrivial
solutions of the equation $p\omega_1+q\omega_2+r\omega_{3R}=0$, where $p$, $q$,
and $r$ are integers, $\omega_1$ and $\omega_2$ are the forcing frequencies,
and $\omega_{3R}$ is the resonant response, and can be written compactly in the
form $[p,q,r]$. Combination tones are three-frequency resonances of
the restricted class $[p,q,1]$. This is the only type of response possible from
a passive nonlinearity, whereas a dynamical system such as a forced oscillator
is an active nonlinearity with at least one intrinsic frequency, and can
exhibit the full panoply of three-frequency resonances, which include
subharmonics of combination tones. Three-frequency resonances obey
hierarchical ordering properties very similar to those governing two-frequency
resonances in periodically forced systems. In the interval
$(\omega_2/p,\omega_1/q)$, we may define a generalized Farey sum between any
pair of adjacents as $a_1/c\oplus a_2/d=(a_1+a_2)/(c+d)$. The daughter
three-frequency resonance characterized by the generalized mediant is the 
widest between its parents characterized by the adjacents \cite{3freq}.
Thus three-frequency resonances are ordered very similarly to their
counterparts in two-frequency systems, and form their own devil's staircase;
Fig.~\ref{3f_devils_staircase}.
\vspace{1cm}

\end{document}